\documentclass[aps,preprint,amsmath,amssymb,showpacs,amsfonts]{revtex4}
\usepackage{epsfig}
\usepackage{graphicx}
\usepackage{dcolumn}
\usepackage{bm}

\newcommand{\del}{\partial}

\newcommand{\dd}[3]{\left(\frac{\del #1}{\del #2}\right)_{\! #3}} 

\begin{document}

\title{Relativistic Hydrodynamics with General Anomalous Charges}

\author{Yasha Neiman}
\author{Yaron Oz}
\affiliation{Raymond and Beverly Sackler School of Physics and Astronomy, Tel-Aviv University, Tel-Aviv 69978, Israel}

\date{\today}

\begin{abstract}

We consider the hydrodynamic regime of gauge theories with general triangle anomalies, where the participating currents may be global or gauged, abelian or non-abelian.
We generalize the argument of arXiv:0906.5044, and construct at the viscous order the stress-energy tensor, the charge currents and the entropy current.

\end{abstract}
\maketitle

\section{Introduction and Summary}

In this letter we will consider the hydrodynamic regime of microscopic quantum gauge theories with triangle anomalies. The 
field theory currents can be global or gauged, abelian or non-abelian.  
We will label this full set of currents with the indices $(\alpha,\beta,\dots)$, and will limit ourselves to symmetries which are not spontaneously broken.

Some of the field theory charges will be accessible at the energy scale defined by the temperature. Such charges and their chemical potentials may take part in the hydrodynamic description, as part of the local equilibrium parameters. There are possible obstructions to this, e.g. a gauged charge may be screened. Also, to be part of the thermodynamics a charge must be conserved with sufficient accuracy within each local equilibration region; thus, a global $U(1)$ charge will not be part of the fluid description if its conservation is too disrupted by anomalies with gauged currents. For the bulk of the discussion, we will assume that the conservation of hydrodynamic charges is not violated at all. However, during the derivation we \emph{will} violate the conservation of charges using fictitious external fields. The more general case of slight violation will be discussed in section \ref{sec:non-conservation}.

Gauge fields may also enter the fluid description by assuming nonzero average values over macroscopic distances. This possibility upgrades the discussion from pure hydrodynamics to electro/magneto-hydrodynamics. If a gauge field enters the description, so will the current that it induces; the associated charge density may also appear as an independent parameter, or it may be screened.

For non-abelian gauge fields and their associated charges, a complication arises. If a charge density is to be part of the thermal parameters, it must be additive and conserved (in the ordinary, not in the gauge-covariant sense) within each local equilibration region. In other words, within each equilibration region the charge must be approximately abelian. This imposes a condition on the fluctuating microscopic component of the associated gauge field. The same condition must hold for the gauge field itself to enter a fluid description: without a consistent gauge frame within each equilibration region, the gauge field cannot have a smooth, well-defined average value. The technical consequences of this condition are that gauge-covariant derivatives commute with thermal averaging, and that the averaged field strength $F^a_{\mu\nu}$ can be derived from the averaged potential $A^a_\mu$.

Our aim is to find the allowed leading viscous order constitutive relations of the field theory hydrodynamics. In order to keep track of orders of magnitude, we introduce a formal small parameter $\varepsilon$. The ``order'' of a quantity will refer to the power of $\varepsilon$ involved. We will take all the gradients and field strengths to be of order $\sim \varepsilon$. The background metric $g_{\mu\nu}$ may be curved on the length scale of the hydrodynamic gradients, i.e. its Riemann tensor is of order $O(\varepsilon^2)$.

The zeroth-order constitutive relations for the stress-energy density, the charge currents and the entropy current have the standard ideal-fluid form:
\begin{align}
 \begin{split}
   T^{(0)\nu}_\mu &= \sqrt{-g}(\epsilon u_\mu u^\nu + p P_\mu^\nu)  \ , \\
   J^{(0)\mu}_a &= \sqrt{-g}n_a u^\mu  \ , \\
   s^{(0)\mu} &= \sqrt{-g}s u^\mu \ , \label{eq:ideal_constitutive}
 \end{split}
\end{align}
where $u^\mu$ is the energy velocity, $u_\mu u^\mu = -1$, $\epsilon$ is the energy density, $p$ is the pressure, $n_a$ is the charge density and $s$ is the entropy density. $P_\mu^\nu = \delta_\mu^\nu + u_\mu u^\nu$ is the projector orthogonal to $u^\mu$. The indices $(a,b,\dots)$ enumerate the currents which participate in the hydrodynamics (whether through a charge density or a gauge field). For screened charges, we set $n_a = 0$. $T_\mu^\nu$ is the ``intrinsic'' stress-energy density, i.e. it includes the contribution from the microscopic component of the gauge fields, but not the contribution from the macroscopic gauge fields.

Our result for the first-order constitutive relations is a direct generalization of \cite{Son:2009tf}. We will find (in the Landau frame):
\begin{align}
 \begin{split}
   T^{(1)\nu}_\mu &= -\sqrt{-g}(2\eta\pi_\mu^\nu + \zeta P_\mu^\nu D_\rho u^\rho) \ , \\
   J^{(1)\mu}_a &= \sqrt{-g}\sigma_a{}^b\left(E_b^\mu - T P^{\mu\nu} D_\nu\frac{\mu_b}{T}\right) + \xi_a\omega^\mu + \xi^{(B)}_{ab} B^{b\mu} \ , \\
   s^{(1)\mu} &= -\frac{\mu^a}{T}J_a^{(1)\mu} + \tilde\xi\omega^\mu + \tilde\xi^{(B)}_a B^{a\mu} \ , \label{eq:viscous_constitutive}
 \end{split}
\end{align}
where $T = (\partial\epsilon/\partial s)_{\mathbf{n}}$ is the temperature, $\mu_a = (\partial\epsilon/\partial n^a)_s$ are the chemical potentials (with $\mu_a \equiv 0$ for screened charges), $\eta \geq 0$ is the shear viscosity, $\zeta \geq 0$ is the bulk viscosity, and $\sigma_{ab}$ is the conductivity matrix with $\sigma_{(ab)}$ positive semi-definite. $D_\mu$ is the covariant derivative with respect to the metric and the gauge fields. The derivative of $\mu_a/T$ is taken in the gauge-covariant sense, i.e. $D_\nu\mu_a = \del_\nu\mu_a + f_{abc}A^b_\nu\mu^c$, where $f_{abc}$ are the Lie algebra structure constants. The shear tensor $\pi_{\mu\nu}$ and the vorticity density $\omega^\mu$ are defined by:
\begin{align}
 \begin{split}
   \pi_{\mu\nu} &\equiv P^\rho_\mu P^\sigma_\nu D_{(\rho}u_{\sigma)} - \frac{1}{3}P_{\mu\nu} D_\rho u^\rho \ , \\
   \omega^\mu &\equiv \frac{1}{2}\epsilon^{\mu\nu\rho\sigma} u_\nu \del_\rho u_\sigma \ ,
 \end{split}
\end{align}
where $\epsilon^{\mu\nu\rho\sigma}$ is the metric-independent Levi-Civita density with components $\pm 1$. The electric field $E_a^\mu$ and magnetic field $B_a^\mu$ are defined by:
\begin{align}
  E_a^\mu \equiv F^{\mu\nu}_a u_\nu; \quad B^{a\mu} \equiv \frac{1}{2}\epsilon^{\mu\nu\rho\sigma}u_\nu F^a_{\rho\sigma} \label{eq:E_B} \ ,
\end{align}
where $F^a_{\mu\nu}$ is the macroscopic gauge field strength. For global charges, we set $F^a_{\mu\nu} = 0$.

The ordinary viscous coefficients $\zeta$, $\eta$ and $\sigma_{ab}$ can be arbitrary functions of state. In contrast, the vorticity/magnetic coefficients $\xi_a$, $\xi^{(B)}_{ab}$, $\tilde\xi$ and $\tilde\xi^{(B)}_a$ are almost entirely fixed by the field theory's chiral anomalies:
\begin{align}
 \begin{split}
    \xi_a &= C_{abc}\mu^b\mu^c + 2\beta_a T^2 - \frac{2n_a}{\epsilon + p}\left(\frac{1}{3}C_{bcd}\mu^b\mu^c\mu^d + 2\beta_b\mu^b T^2 + \gamma T^3\right) \ , \\
    \xi^{(B)}_{ab} &= C_{abc}\mu^c - \frac{n_a}{\epsilon + p}\left(\frac{1}{2}C_{bcd}\mu^c\mu^d + \beta_b T^2\right) \ , \\
    \tilde\xi &= \frac{1}{3T}C_{abc}\mu^a\mu^b\mu^c + 2\beta_a\mu^a T + \gamma T^2 \ , \\
    \tilde\xi^{(B)}_a &= \frac{1}{2T}C_{abc}\mu^b\mu^c + \beta_a T \ . \label{eq:xi_results}
 \end{split}
\end{align}
Here, $C_{abc}$ is a symmetric tensor of anomaly coefficients:
\begin{align}
 C_{abc} = \frac{1}{4\pi^2}\operatorname{tr}\{ T_{(a} T_b T_{c)} \}, \label{eq:C}
\end{align}
where $T_a$ are the symmetry generators in the fermion representation. $\beta_a$ and $\gamma$ are numerical constants, which are not constrained by the anomaly.
Note that a nonzero $\gamma$ is possible only for a parity-breaking theory. $\beta_a$, on the other hand, is allowed in parity-conserving theories, as long as it is nonzero only for values of $a$ corresponding to axial charges. Also, the group structure forbids a nonzero $\beta_a$ when $a$ corresponds to a non-abelian charge.

Our derivation generalizes the results of \cite{Son:2009tf} by allowing for gauged and/or non-abelian charges. We also elucidate the precise definition of the currents used in the derivation, leading to the correct value for $C_{abc}$. Finally, we include the additional vorticity coefficients $\beta_a$ and $\gamma$, which were omitted in \cite{Son:2009tf}. The results are consistent with our previous derivation for global non-abelian charges based on a null horizon dynamics in a gravitational dual description \cite{Eling:2010hu}.

The appearance of the vorticity term $\omega^{\mu}$ in the anomalous hydrodynamic current has been first observed in the dual gravitational description of relativistic conformal hydrodynamics \cite{Erdmenger:2008rm,Banerjee:2008th}. Experimental manifestations of the vorticity term in heavy ion collisions have been proposed in  \cite{KerenZur:2010zw,Kharzeev:2010gr}. We note that hydrodynamics with $SU(2)$ global symmetry charges has been considered in a holographic setup in \cite{Torabian:2009qk}.

The paper is organized as follows. In section \ref{sec:currents}, we introduce external fields and carefully define the anomalous currents and their non-conservation equations. In section \ref{sec:constitutive}, we run through the derivation of eqs. \eqref{eq:xi_results}. The derivation closely follows the one in \cite{Son:2009tf}, with gauge-covariant derivatives introduced where necessary. It is essential for the argument that the currents and (non-)conservation equations defined in section \ref{sec:currents} are gauge-covariant, and that $C_{abc}$ assumes a symmetric form. Finally, in section \ref{sec:non-conservation} we consider global currents whose conservation is slightly violated by anomalies with gauge fields.

\section{External fields and definition of the currents} \label{sec:currents}

In order to better exploit the second law of thermodynamics, we will extend the theory by coupling the global hydrodynamic currents to external gauge fields. After this step, there is a gauge field $A^a_\mu$ coupled to each hydrodynamic current. The external gauge fields will enter along with the dynamical fields into the anomalous non-conservation equations. The external field strengths are of order $\sim \varepsilon$, and are approximately homogeneous within each equilibration region; they may vary over the length scales of the hydrodynamic gradients. The covariant derivative $D_\mu$ will be understood to take the external fields into account.

In a theory with anomalies, two subtleties arise with regard to the definition of the currents and the precise form of their non-conservation equations. First, the anomaly diagram may be regularized in different ways, by adding different local counterterms to the action \cite{Bardeen:1969md}. These counterterms are given by local functionals of $A^a_\mu$, and affect the definition of the current. Furthermore, the current may be defined not directly from the variation of the path integral, but again with the addition of some local functional of $A^a_\mu$ \cite{Bardeen:1984pm}. Since there are no anomalies among the dynamically gauged currents, these issues only arise after we add the external gauge fields. What we would like to have is a set of currents that are gauge-covariant with respect to both the dynamical and the external gauge fields. The covariance can then be used to constrain the hydrodynamic constitutive relations. We will now see that there is essentially one such choice of currents, regardless of the regularization.

Let the anomalous diagrams be regularized in any permissible way (i.e. keeping the dynamical gauge symmetries intact). Consider a spacetime region $\Omega$, with boundary conditions on $\del\Omega$ corresponding to our hydrodynamic state. Let us perform the path integral within $\Omega$ with these boundary conditions, without integrating over the gauge fields:
\begin{align}
 W[g_{\mu\nu}, A^\alpha_\mu] = -i\ln\int{\mathcal{D}\psi\mathcal{D}\bar\psi\mathcal{D}\phi\, e^{iS[g_{\mu\nu}, A^\alpha_\mu, \psi, \bar\psi, \phi]}} \ ,
\end{align}
where $\psi$ are the fermion fields responsible for the anomaly, and $\phi$ are any other fields present in the theory. Our notation shows the dependence of $W$ on the values of $g_{\mu\nu}$ and $A^\alpha_\mu$ within $\Omega$, and suppresses its dependence on the boundary conditions on $\del\Omega$. The gauge fields $A^\alpha_\mu$ include both the dynamical fields, whether or not they enter the fluid description, and the external fields.

We now wish to derive the currents and the stress-energy density by varying $W$ with respect to $A^\alpha_\mu$ and $g_{\mu\nu}$, respectively. However, in the hydrodynamic context we do not want the macroscopic currents to contain the $D_\nu F^{\mu\nu}$ contribution derived from the kinetic term of the dynamical gauge fields; also, we do
not want the stress-energy density to contain the stress-energy of the macroscopic piece of the dynamical gauge fields. Therefore, we first subtract from $W$ the kinetic term of the macroscopic gauge fields:
\begin{align}
 W' = W + \frac{1}{4}\int d^4x \sqrt{-g} \left<F^{\bar a}_{\mu\nu}\right> \left<F_{\bar a}^{\mu\nu}\right> \ ,
\end{align}
where the index $\bar a$ runs over the dynamical gauge fields, and the brackets denote averaging over the local equilibration region. This adjustment is gauge-invariant, so the gauge transformation properties of $W'$ are the same as for $W$.

We now define the ``consistent'' anomalous currents as
\begin{equation}
j_\alpha^\mu = \delta W'/\delta A^\alpha_\mu \ .
\end{equation}
The divergence $D_\mu j_\alpha^\mu$ generates gauge transformations inside $\Omega$. The $j_\alpha^\mu$ are not gauge-covariant, but they can be made covariant by the addition of a local functional of $A^\alpha_\mu$. This is best explained in two steps. First, let $G$ be the counterterm to the action which would be necessary to pass from our chosen regularization of the anomalous diagrams to the standard symmetric one (which would violate the dynamical gauge symmetries). Now, let us add $\delta G/\delta A^\alpha_\mu$ to our currents. The resulting currents have the same transformation properties and satisfy the same non-conservation equations as the ``consistent'' currents under the symmetric regularization. These are still not gauge-covariant. However, as shown in \cite{Bardeen:1984pm}, we can add to them an additional local functional $X^\mu_\alpha[A^\alpha_\mu]$, which brings the currents into a gauge-covariant form. We denote the new covariant currents by $J_\alpha^\mu$. They satisfy the non-conservation equation:
\begin{align}
 D_\mu J_\alpha^\mu = \frac{1}{8}C_{\alpha\beta\gamma}\epsilon^{\mu\nu\rho\sigma}F^\beta_{\mu\nu}F^\gamma_{\rho\sigma} \ , \label{eq:div_J_tot}
\end{align}
where $C_{\alpha\beta\gamma}$ is given by \eqref{eq:C}, with the indices $\alpha\beta\gamma$ of the full symmetry algebra instead of $abc$. We stress that eq. \eqref{eq:div_J_tot} depends only on the matter content of the theory, and not on the choice of regularization. Also, in this context there is no conflict between the symmetric form of \eqref{eq:div_J_tot} and the exact gauge invariance of the dynamically gauged currents.

There are essentially no other choices of a gauge-covariant current: the only gauge-covariant functional of $A^\alpha_\mu$ with the right dimension that can be added to $J_\alpha^\mu$ is $D_\nu F_\alpha^{\mu\nu}$. For the hydrodynamic currents, such a term was intentionally excluded. In any case, the divergence of this term vanishes identically, so it would not change eq. \eqref{eq:div_J_tot}. We also note that if one is interested in the constitutive relations for some other choice of currents, one should simply take the results \eqref{eq:ideal_constitutive} and \eqref{eq:viscous_constitutive} for $J_a^\mu$ and add to them the appropriate gauge-field functional. In principle, we should have another contribution to $D_\mu J_\alpha^\mu$ from anomalies with gravity. However, these are proportional to the square of the Riemann tensor, which is $O(\varepsilon^4)$. The gravitational contribution is therefore negligible at the relevant hydrodynamic order.

We now turn to the stress-energy density. So far, we have not performed the path integral over the dynamical gauge fields. Let us now integrate over all the gauge fields except for the macroscopic piece of the fields $A^a_\mu$ that participates in the hydrodynamics. This yields a new functional $W'[g_{\mu\nu},\left<A^a_\mu\right>]$. The stress-energy density is then defined as $T^{\mu\nu} = \delta W'/\delta g_{\mu\nu}$. This is a gauge-invariant quantity, since variations $\delta g_{\mu\nu}$ of the metric commute with gauge transformations $\delta\Lambda^a$, and the anomaly $\delta W'/\delta\Lambda^a$ is a metric-independent functional of gauge fields. From the transformation law under diffeomorphisms inside $\Omega$, we obtain the conservation law:
\begin{align}
 D_\nu T^\nu_\mu = \left<F^a_{\mu\nu}\right> \left<j_a^\nu\right> - \left<A^a_\mu\right> D_\nu \left<j_a^\nu\right> \ , \label{eq:div_T_raw}
\end{align}
where the covariant derivative on the RHS is defined in terms of $\left<A^a_\mu\right>$. Eq. \eqref{eq:div_T_raw} can be written in a manifestly gauge-covariant way as:
\begin{align}
 D_\nu T^\nu_\mu = \left<F^a_{\mu\nu}\right> \left<J_a^\nu\right> \ . \label{eq:div_T}
\end{align}
To see that the two equations are equivalent, we note that their difference is a functional of gauge fields alone (where the gauge fields that were already integrated out are understood as operators). Also, the difference must be gauge-invariant, since the LHS of both equations and the RHS of the second one are manifestly gauge-invariant. It must also be of mass dimension five. Finally, recalling the form of $D_\nu j_a^\nu$ and $J_a^\nu - j_a^\nu$, we see that the lowest power with which gauge fields appear can in the difference is three. There is no functional satisfying these conditions, so the difference between the two equations must vanish.

It remains to complete the path integration, and to write the average of eq. \eqref{eq:div_J_tot} for the hydrodynamic currents:
\begin{align}
  D_\mu \left<J_a^\mu\right> &= \frac{1}{8}C_{a\beta\gamma}\epsilon^{\mu\nu\rho\sigma}\left<F^\beta_{\mu\nu} F^\gamma_{\rho\sigma}\right> \ . \label{eq:div_J_raw}
\end{align}
On the LHS of this equation, the covariant derivative $D_\mu$ commutes with the averaging brackets, due to the weakness condition on the microscopic component of the dynamical gauge fields which participate in the hydrodynamics. By assumption (which will be relaxed in section \ref{sec:non-conservation}), the RHS vanishes in the absence of external fields. In other words, it receives no contribution from products $\epsilon^{\mu\nu\rho\sigma}F^\beta_{\mu\nu} F^\gamma_{\rho\sigma}$ of two \emph{dynamical} field strengths. Now, the external gauge fields do not have a microscopic component. Therefore, we can pull them out of the averaging brackets, giving:
\begin{align}
  D_\mu \left<J_a^\mu\right> &= \frac{1}{8}C_{abc}\epsilon^{\mu\nu\rho\sigma}\left<F^b_{\mu\nu}\right>\left<F^c_{\rho\sigma}\right> \ . \label{eq:div_J}
\end{align}
We've reduced the summed-over indices into the subspace of hydrodynamic currents, since only the gauge fields associated with such currents have a non-vanishing macroscopic component.

\section{Constitutive relations} \label{sec:constitutive}

The zeroth-order hydrodynamic constitutive relations are given by \eqref{eq:ideal_constitutive}. We will now use the non-conservation equations \eqref{eq:div_T}, \eqref{eq:div_J} and the second law of thermodynamics to derive the possible form of the first-order constitutive relations. From now on we will drop the averaging brackets, understanding that averaged quantities are always implied. Using the definitions \eqref{eq:E_B}, we rewrite the relevant components of eqs. \eqref{eq:div_T}, \eqref{eq:div_J} as:
\begin{align}
 \begin{split}
   u^\mu D_\nu T^\nu_\mu &= E^a_\mu J_a^\mu  \ , \\
   D_\mu J_a^\mu &= C_{abc}E^b_\mu B^{c\mu} \ . \label{eq:hydro_conservation}
 \end{split}
\end{align}

We fix the velocity $u^\mu$ as the unit timelike eigenvector of $T_\mu^\nu$, the energy density $\epsilon$ as the eigenvalue of $T_\mu^\nu/\sqrt{-g}$ corresponding to $u^\mu$, and the charge density as $n_a = u_\mu J_a^\mu/\sqrt{-g}$. Using gauge covariance to constrain the possible contributions, the most general first-order corrections to the constitutive relations read:
\begin{align}
  \begin{split}
    T^{(1)\nu}_\mu ={}& -\sqrt{-h}(2\eta\pi_\mu^\nu + \zeta P_\mu^\nu D_\rho u^\rho) \ , \\
    J^{(1)\mu}_a ={}& \sqrt{-h}\left(\chi_a P^{\mu\nu}\del_\nu p - T\sigma_a{}^b P^{\mu\nu} D_\nu\frac{\mu_b}{T} + \sigma^{(E)}_{ab} E^{b\mu}\right)
      + \xi_a\omega^\mu + \xi^{(B)}_{ab} B^{b\mu} \ , \\
    s^{(1)\mu} ={}& -\frac{\mu^a}{T}J_a^{(1)\mu} + \sqrt{-h}\left(\tilde\zeta u^\mu D_\nu u^\nu + \tilde\chi P^{\mu\nu}\del_\nu p + \tilde\sigma^a P^{\mu\nu} D_\nu\frac{\mu_a}{T}
      + \tilde\sigma^{(E)}_a E^{a\mu}\right) \\
      &+ \tilde\xi\omega^\mu + \tilde\xi^{(B)}_a B^{a\mu} \ .
  \end{split}
\end{align}
The various coefficient functions are constrained by the non-negativity of the entropy production rate. Using eqs. \eqref{eq:hydro_conservation}, the ideal fluid equations and the first law of thermodynamics, the entropy production rate can be written as:
\begin{align}
  \begin{split}
    \del_\mu s^\mu ={}& \frac{1}{T}\left(-T^{(1)\mu\nu} D_\mu u_\nu + J^{(1)a\mu}\left(E_{a\mu} - T D_\mu\frac{\mu_a}{T}\right) - C_{abc}\mu^a E^b_\mu B^{c\mu}\right) \\
      &+ \del_\mu\left(s^{(1)\mu} + \frac{\mu_a}{T}J^{(1)a\mu}\right) \ . \label{eq:div_s}
  \end{split}
\end{align}

The different terms in \eqref{eq:div_s} divide into those with a factor of $\epsilon^{\mu\nu\rho\sigma}$ (via $\omega^\mu$ or $B_a^\mu$) and those without. For the terms without $\epsilon^{\mu\nu\rho\sigma}$, a standard exercise shows that we must  have $\tilde\chi = \tilde\zeta = 0$, $\chi_a = \tilde\sigma_a = \tilde\sigma^{(E)}_a = 0$, $\sigma^{(E)}_{ab} = \sigma_{ab}$, $\eta \geq 0$, $\zeta \geq 0$ and a positive semi-definite $\sigma_{(ab)}$.

Let us now turn to the $\epsilon^{\mu\nu\rho\sigma}$-terms, closely following the derivation of \cite{Son:2009tf}. The only way for these terms to be non-negative is to vanish identically. The following consequences of the ideal fluid equations are useful:
\begin{align}
 \begin{split}
   \del_\mu\omega^\mu &= 2a_\mu\omega^\mu = -\frac{2}{\epsilon + p}\omega^\mu(\del_\mu p - n_a E^a_\mu) \ , \\
   D_\mu B^\mu_a &= -2\omega^\mu E_{a\mu} + a_\mu B_a^\mu = -2\omega^\mu E_{a\mu} - \frac{1}{\epsilon + p}B_a^\mu(\del_\mu p - n_b E^b_\mu) \ .
 \end{split}
\end{align}

We choose $(p,\mu_i/T)$ as the independent thermodynamic parameters on which the coefficients $(\xi_a, \xi^{(B)}_{ab}, \tilde\xi, \tilde\xi^{(B)}_a)$ may depend. We use the indices $i,j,k$ to enumerate the hydrodynamic charges which are not screened, and can therefore assume nonzero values. For the gradients of $\tilde\xi$ and $\tilde\xi^{(B)}_a$ in \eqref{eq:div_s}, we use the chain rule for the covariant derivative:
\begin{align}
  D_\mu\mathbf\alpha = \left(\frac{\del\mathbf\alpha}{\del p}\right)_{\mu_i/T} \del_\mu p + \left(\frac{\del\mathbf\alpha}{\del(\mu_i/T)}\right)_p D_\mu\frac{\mu_i}{T} \ ,
\end{align}
where $\mathbf\alpha$ is any thermodynamic function of state, with arbitrary charge indices.

The condition on the $\epsilon^{\mu\nu\rho\sigma}$-terms in \eqref{eq:div_s} can now be written as:
\begin{align}
  \begin{split}
    & \left(\frac{\del\tilde\xi}{\del p} - \frac{2\tilde\xi}{\epsilon + p}\right)\del_\mu p\,\omega^\mu
    + \left(\frac{\del\tilde\xi}{\del(\mu^i/T)} - \xi_i\right) D_\mu\frac{\mu^i}{T}\omega^\mu
    + \left(\frac{\xi_a}{T} + \frac{2n_a\tilde\xi}{\epsilon + p} - 2\tilde\xi^{(B)}_a\right) E^a_\mu\omega^\mu \\
    &{}+ \left(\frac{\del\tilde\xi^{(B)}_a}{\del p} - \frac{\tilde\xi^{(B)}_a}{\epsilon + p}\right) \del_\mu p\,B^{a\mu}
    + \left(\frac{\del\tilde\xi^{(B)}_b}{\del(\mu^i/T)} - \xi^{(B)}_{ib}\right) D_\mu\frac{\mu^i}{T} B^{b\mu} \\
    &{}+ \left(\frac{1}{T}\xi^{(B)}_{ab} - \frac{1}{T}C_{iab}\mu^i + \frac{n_a\tilde\xi^{(B)}_b}{\epsilon + p}\right) E^a_\mu B^{b\mu} = 0 \ .
  \end{split}
\end{align}
At a point, the second-order terms on the right of each set of parentheses are independent. Therefore, all the coefficients must vanish separately:
\begin{align}
  & \frac{\del\tilde\xi}{\del p} = \frac{2\tilde\xi}{\epsilon + p}; &
  & \frac{\del\tilde\xi^{(B)}_a}{\del p} = \frac{\tilde\xi^{(B)}_a}{\epsilon + p} \label{eq:d_d_p} \\
  & \frac{\del\tilde\xi}{\del(\mu^i/T)} = \xi_i; &
  & \frac{\del\tilde\xi^{(B)}_b}{\del(\mu^i/T)} = \xi^{(B)}_{ib} \label{eq:d_d_mu} \\
  & \frac{\xi_i}{T} + \frac{2n_i\tilde\xi}{\epsilon + p} = 2\tilde\xi^{(B)}_i; &
  & \frac{1}{T}\xi^{(B)}_{ib} + \frac{n_i\tilde\xi^{(B)}_b}{\epsilon + p} = \frac{1}{T}C_{jib}\mu^j \label{eq:equals_C_i} \\
  & \frac{\xi_q}{T} = 2\tilde\xi^{(B)}_q; &
  & \frac{1}{T}\xi^{(B)}_{qb} = \frac{1}{T}C_{iqb}\mu^i \label{eq:equals_C_q}
\end{align}
where the index $q$ runs over the screened charges, for which $n_q = 0$. To solve these equations, we use the thermodynamic identities:
\begin{align}
  \dd{T}{p}{\mu_i/T} = \frac{T}{\epsilon + p}; \quad \dd{T}{(\mu_i/T)}{p} = -\frac{n^i T^2}{\epsilon + p} \ .
\end{align}
Eq. \eqref{eq:d_d_p} then implies that $\tilde\xi/T^2$ and $\tilde\xi^{(B)}_a/T$ are functions of $\mu_i/T$ only. Using eq. \eqref{eq:d_d_mu} to eliminate $\xi_i$ and $\xi^{(B)}_{ib}$, eq. \eqref{eq:equals_C_i} then gives:
\begin{align}
  \frac{\del(\tilde\xi/T^2)}{\del(\mu^i/T)} &= \frac{2}{T}\tilde\xi^{(B)}_i \ , &
  \frac{\del(\tilde\xi^{(B)}_b/T)}{\del(\mu^i/T)} &= \frac{1}{T}C_{jib}\mu^j  \ . \label{eq:tilde_xi_C}
\end{align}

For eq. \eqref{eq:tilde_xi_C} to be integrable, we must have the symmetry conditions $C_{ijb} = C_{(ij)b}$ and $C_{ijk}$ = $C_{(ijk)}$. As we saw in section \ref{sec:currents}, this is guaranteed by our construction of gauge-covariant currents. Integrating and taking into account eq. \eqref{eq:equals_C_q}, we get the result \eqref{eq:xi_results}. The numbers $\beta_a$ and $\gamma$ arise as integration constants. Taking away the external fields, we obtain the constitutive relations of the physical fluid.

\section{Non-conservation due to dynamical gauge fields} \label{sec:non-conservation}

In the above discussion, we assumed that in the absence of external fields, the hydrodynamic currents are exactly conserved. This assumption can be relaxed. To be part of the fluid description, a global charge only needs to be conserved approximately, so that its production rate is negligible within each local equilibration region. However, on the scale of hydrodynamic gradients, the conservation of the charge may be violated, as it was in sections \ref{sec:currents}-\ref{sec:constitutive} due to anomalies with external gauge fields. The same kind of ``soft'' non-conservation may take place in the physical fluid with no external fields, due to anomalies with the dynamical fields. Taking into account both kinds of anomalies, the averaged non-conservation equation \eqref{eq:div_J} takes the form:
\begin{align}
 \begin{split}
   D_\mu \left<J_a^\mu\right> &= \frac{1}{8}C_{abc}\epsilon^{\mu\nu\rho\sigma}\left<F^b_{\mu\nu}\right>\left<F^c_{\rho\sigma}\right> + \Phi_a \ , \\
   \Phi_a &\equiv \frac{1}{8}C_{a\beta\gamma}\epsilon^{\mu\nu\rho\sigma}\left<F^\beta_{\mu\nu} F^\gamma_{\rho\sigma}\right>_{micro} \ , \label{eq:div_J_Phi}
 \end{split}
\end{align}
where $\Phi_a$ is the charge production rate due to anomalies with the microscopic component of the dynamical gauge fields. Due to the group structure, $\Phi_a$ can be nonzero only for global $U(1)$ charges. Like every quantity in the hydrodynamic regime, $\Phi_a$ will be a local functional of the thermodynamic parameters and the macroscopic gauge fields.

The $\Phi_a$-term provides a correction to the hydrodynamic charge production rate in \eqref{eq:hydro_conservation}:
\begin{align}
  D_\mu J_a^\mu &= C_{abc}E^b_\mu B^{c\mu} + \Phi_a \ .
\end{align}
This propagates into the entropy production rate \eqref{eq:div_s}, which becomes:
\begin{align}
  \begin{split}
    \del_\mu s^\mu ={}& \frac{1}{T}\left(-T^{(1)\mu\nu} D_\mu u_\nu + J^{(1)a\mu}\left(E_{a\mu} - T D_\mu\frac{\mu_a}{T}\right) - C_{abc}\mu^a E^b_\mu B^{c\mu} - \mu_a\Phi^a\right) \\
      &+ \del_\mu\left(s^{(1)\mu} + \frac{\mu_a}{T}J^{(1)a\mu}\right) \ . \label{eq:div_s_Phi}
  \end{split}
\end{align}

In powers of the small parameter $\varepsilon$, we are interested in $\Phi_a$ up to second order. A zeroth-order contribution to $\Phi_a$ must be suppressed by some small dimensionless factor, in order to maintain the approximate conservation of charge within each local equilibration region. Such a term does not mix with any other term in \eqref{eq:div_s_Phi}. Therefore, its contribution to the entropy production must be always non-negative, i.e. $\mu_a\Phi^{(0)a} \leq 0$. A first-order contribution $\Phi_a^{(1)}$ also cannot mix with any other term in \eqref{eq:div_s_Phi}. Such a contribution is then ruled out by the second law of thermodynamics, since it must be proportional to $D_\mu u^\mu$, with no way to constrain its sign. Finally, a second-order term $\Phi_a^{(2)}$ is not ruled out by the second law, and it \emph{will} mix with the other terms in \eqref{eq:div_s_Phi}. In fact, such an arbitrary second-order contribution would affect the entire set of constraints on the transport coefficients, and invalidate our conclusions. Thus, the only situation in which we can make a clear statement is when $\Phi_a^{(2)}$ can be neglected, and only $\Phi_a^{(0)}$ must be taken into account. In fact, this is a reasonable assumption: this is what we get if the same suppression factor present in $\Phi_a^{(0)}$ also affects $\Phi_a^{(2)}$, in addition to the suppression from small derivatives.

In conclusion, we assume that $\Phi_a$ is given by some thermodynamic function of state, without derivatives or gauge field factors. This function must be suppressed by some small number, and satisfy $\mu_a\Phi^a \leq 0$. Our results for the transport coefficients remain unchanged, and the entropy production rate reads:
\begin{align}
  \begin{split}
    \del_\mu s^\mu ={}& \frac{\sqrt{-h}}{T}\left(2\eta\pi_{\mu\nu}\pi^{\mu\nu} + 3\zeta(D_\mu u^\mu)^2
       + \sigma^{ab}\left(E_{a\mu} - TD_\mu\frac{\mu_a}{T}\right)\left(E_b^\mu - TD^\mu\frac{\mu_b}{T}\right)\right) \\
      & - \frac{1}{T}\mu_a\Phi^a \ .
  \end{split}
\end{align}

\section*{Acknowledgements}		

We would like to thank D. T. Son for an e-mail exchange. The work is supported in part by the Israeli Science Foundation center of excellence, by the Deutsch-Israelische Projektkooperation (DIP), by the US-Israel Binational Science Foundation (BSF), and by the German-Israeli Foundation (GIF).

\end{document}